\documentclass[aps,pre,twocolumn,amsmath,amssymb,superscriptaddress,floatfix]{revtex4} 

\usepackage{graphicx}
\usepackage{bm}
\usepackage[usenames]{color}
\bibstyle{apsrev.bib}

\newcommand{\be}{\begin{equation}}
\newcommand{\ee}{\end{equation}}
\newcommand{\beqn}{\begin{eqnarray}}
\newcommand{\eeqn}{\end{eqnarray}}

\begin{document}

\title{Mixed-order phase transition of the contact process near multiple junctions}
\author{R\'obert Juh\'asz}
\email{juhasz.robert@wigner.mta.hu}
\affiliation{Wigner Research Centre for Physics, Institute for Solid State
Physics and Optics, H-1525 Budapest, P.O.Box 49, Hungary}
\author{Ferenc Igl\'oi}
\email{igloi.ferenc@wigner.mta.hu}
\affiliation{Wigner Research Centre for Physics, Institute for Solid State
Physics and Optics, H-1525 Budapest, P.O.Box 49, Hungary}
\affiliation{Institute of Theoretical Physics, Szeged University, H-6720 Szeged,
Hungary}
\date{\today}

\begin{abstract}
We have studied the phase transition of the contact process near a multiple junction of $M$ semi-infinite chains by Monte Carlo simulations. 
As opposed to the continuous transitions of the translationally invariant ($M=2$) and semi-infinite ($M=1$) system, the local order parameter is found to be discontinuous for $M>2$.  
Furthermore, the temporal correlation length diverges algebraically as the critical point is approached, but with different exponents on the two sides of the transition. In the active phase, the estimate is compatible with the bulk value, while in the inactive phase it exceeds the bulk value and increases with $M$. The unusual local critical behavior is explained by a scaling theory with an irrelevant variable, which becomes dangerous in the inactive phase.
Quenched spatial disorder is found to make the transition continuous in agreement with earlier renormalization group results.  
\end{abstract}

\maketitle

\section{Introduction}

The contact process \cite{cp,liggett} is a stochastic lattice model, which is interpreted most frequently as a simple model of epidemic spreading or population dynamics. It consists of two concurring processes: the activation of nearest neighbor inactive sites and spontaneous deactivation. Tuning the relative rate of these two, the model undergoes a nonequilibrium phase transition from an active, fluctuating phase to a non-fluctuating (absorbing) one 
\cite{md,odor,hhl}.
The transition is continuous in any dimension and falls into the robust universality class of directed percolation. Although the model is not exactly soluble even on regular lattices, the critical exponents and the location of the critical point is known at high precision by series expansions \cite{jd,jensen}.  

The phase transition of the contact process, being a prototypical model of absorbing phase transitions, has been studied in the past beyond translational invariant regular lattices under various circumstances such as near surfaces \cite{jss,egjt,lfh,jensen_s,fhl,hfl} or a single defect site \cite{bh} that break translational invariance,
with quenched spatial \cite{moreira,hiv,vojta_rev} or temporal disorder \cite{jensen_td,vazquez,vh}, long-range interactions \cite{janssen,howard}, on fractals \cite{dsh} and different kinds of complex networks \cite{c_ps,munoz} etc. 
In all the above cases, the transition, although characterized by different critical exponents from those of directed percolation, is observed to remain continuous. 
In general, discontinuous phase transitions in low dimensional fluctuating systems are rare as fluctuations, which destabilize the ordered state, are pronounced in low dimensions. Particularly, in one dimensional fluctuating systems, first-order phase transitions are conjectured to be impossible provided there are no long-range interactions, additional conservation laws, macroscopic currents, or special boundary conditions \cite{hhl}.    
In this work, we shall demonstrate by numerical simulations that a suitable topology of the underlying network is able to induce a discontinuous local transition even with a simple dynamics such as the contact process, which does not display a first-order transition on translationally invariant lattices in any high dimensions.
To be concrete, we shall consider a multiple junction, which consists of $M$ semi-infinite one-dimensional lattices connected to a common central site. This type of star-like  geometry has already been investigated for different type of problems: for the classical and quantum Ising models \cite{itb,cardy,it,tsvelik,qcpmj,monthus,grassberger}, wetting \cite{indekeu}, self-avoiding random walks, percolation \cite{grassberger} etc. For the Ising model, the limit $M \to 0$ corresponds to the problem of random boundary field \cite{itb,cardy,pbti}.
The multiple junction could be a simplified model for describing the behavior of the contact process on complex networks composed of long, one-dimensional segments and rarely located junction points, in the vicinity of junctions. 
This model additionally with quenched disorder has been studied earlier by one of us by means of a renormalization group (RG) method suitable for disordered systems \cite{qcpmj}. Here, we will present numerical results for the clean system, which shows a much different critical behavior and also compare results of the RG method with simulation results of the disordered model. 
The special case $M=2$ of the model is simply the translationally invariant one-dimensional contact process, while $M=1$ is a single semi-infinite chain with an absorbing wall. In the latter case, the surface order parameter $P_1$ is reduced compared to the bulk \cite{jss,egjt,lfh,fhl,hfl} and, approaching the critical point from the active phase it vanishes with the control parameter $\Delta$ as $P_1\sim \Delta^{\beta_1}$, where $\beta_1$ is greater than the corresponding bulk exponent, see in Table \ref{table_exponents}. This kind of behavior is analogous to the ordinary surface transition of equilibrium systems \cite{binder,diehl,ipt,pleimling}. 
The exponent $\nu_{\parallel,1}$ describing the divergence of the temporal correlation length near the surface through $\xi_{\parallel,1}\sim |\Delta|^{-\nu_{\parallel,1}}$, however, holds to be equal to the bulk value, $\nu_{\parallel,1}=\nu_{\parallel}$. The bulk and surface critical exponents of the one-dimensional contact process are summarized in Table \ref{table_exponents}. 
%%%%%%%%%%%%%%%%%%%%%%%%%%%%%%%%%%%%%%%%%%%%%%%%%%%%%%%%%%%%%%%%%%%%%%%%
%%%%%%%%%%%%%%%%%%%%%%%%%%%%%%%%%%%%%%%%%%%%%%%%%%%%%%%%%%%%%%%%%%%
\begin{table}[h]
\begin{center}
\begin{tabular}{|c|l|}
\hline  $\lambda_c$  &  $3.29785(2)$  \cite{jd} \\
\hline  $\beta$    &  $0.276486(8)$ \cite{jensen} \\
\hline  $\beta_1$   &  $0.73371(2)$ \cite{jensen_s} \\
\hline  $\nu_{\parallel}=\nu_{\parallel,1}$    &  $1.733847(6)$ \cite{jensen} \\
\hline  $\nu_{\perp}$   &  $1.096854(4)$ \cite{jensen} \\
\hline
\end{tabular}
\end{center}
\caption{\label{table_exponents} The critical activation rate, as well as bulk and surface critical exponents of the one-dimensional contact process.}
\end{table}
%%%%%%%%%%%%%%%%%%%%%%%%%%%%%%%%%%%%%%%%%%%%%%%%%%%%%%%%%%%%%%%%%%%%%%

This kind of surface critical behavior of the semi-infinite system might suggest a similar scenario for $M>2$, i.e. a continuous surface transition, characterized by an order-parameter exponent $\beta_M$, which is now less than $\beta_2\equiv\beta$ due to the enhanced ordering tendency at the junction, and a temporal correlation length exponent, which is independent of $M$, $\nu_{\parallel,M}=\nu_{\parallel}$. Surprisingly, it turns out that, for $M>2$, the surface order parameter is finite in the bulk critical point and thus displays a jump when the bulk critical point is crossed. This means that the local order-parameter exponent is formally $\beta_M=0$. Nonetheless, spatial and temporal correlation lengths diverge like at continuous transitions. Such kind of transitions are termed as mixed-order transitions and appear in various areas, such as in the classical Ising chain with long-range interactions \cite{anderson1969exact,thouless1969long,dyson1971ising,cardy1981one,aizenman1988discontinuity,slurink1983roughening,bar2014mixed} and in other models, too \cite{fisher1966effect,blossey1995diverging,fisher1984walks,gross1985mean,causo,dorogovtsev,goltsev,baxter,toninelli2006jamming,toninelli2007toninelli,schwarz2006onset,liu2012core,liu2012extraordinary,zia2012extraordinary,bizhani2012discontinuous,ghanbarnejad1,ghanbarnejad2,sheinman2014discontinuous,bar2014mixed1,kji,RBPM}.
A further interesting point is that the temporal correlation-length exponents on the two sides of the transition point are found to be asymmetric;  $\nu_{\parallel,M}$ in the inactive phase departs from $\nu_{\parallel}$ for $M>2$, while the exponent $\nu_{\parallel,M}'$ in the active phase is found to be independent of $M$. 

The rest of the paper is organized as follows. In section \ref{sec:model}, the model is defined and details of the simulation are given. In section \ref{sec:numerical}, the results of Monte Carlo simulations are presented, which are explained in terms of a scaling theory in section \ref{sec:scaling}. The effect of quenched disorder is considered in section \ref{sec:disorder}. Finally, results are discussed in section \ref{sec:discussion}.

%%%%%%%%%%%%%%%%%%%%%%%%%%%%%%%%%%%%%%%%%%%%%%%%%%%%%%%%%%%%%%%%%%%%%%%%%%%%
\section{The model}
\label{sec:model} 

\begin{figure}[h]
\includegraphics[width=7cm]{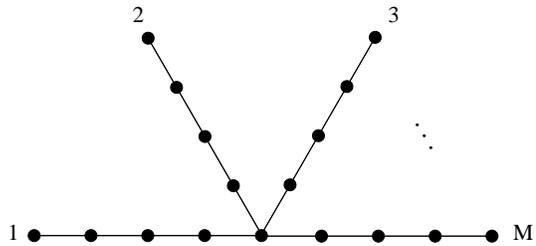}  
%junctionb.eps
\caption{\label{junction} A star-like network composed of $M$ one-dimensional chains sharing a common site.     
}
\end{figure}

The star-like network we considered is composed of $M$ one-dimensional lattices of $L$ sites, in such a way that one of the end sites of each chain (arm) is connected to a central site, see Fig. \ref{junction}. The coordination number of the central site is thus $M$, for the other end sites of the arms it is $1$, while, for all other sites it is $2$. We note that a multiple junction can be defined also in such a way, that the endpoints of the chains at the junction are connected to any other endpoints; nevertheless, the local critical behavior in the two geometries are expected to be identical.
%%%%%%%%%%%%%%%%%%%%%%%%%%%%%%%%%%%%%%%%%%%%%%%%%%%%%%%%%%%%%%%%%%%%%%%%%
%%%%%%%%%%%%%%%%%%%%%%%%%%%%%%%%%%%%%%%%%%%%%%%%%%%%%%%%%%%%%%%%%%%%%%%%

In the contact process, a bimodal variable representing active or inactive states is attached to each sites, and a continuous-time Markov process is defined with two kinds of independent transitions. Active sites either become spontaneously inactive with rate $1$ or activate neighboring inactive sites with a rate $\lambda/n$, where $n$ is the coordination number of the source site. 

In the numerical simulations time was discretized and the process was implemented as follows. A site was chosen randomly from a list of active sites and, with a probability $1/(1+\lambda)$ it was made inactive, or, with a probability $\lambda/(1+\lambda)$ one of its neighbors was chosen randomly (with equal probabilities) and made active, provided it had been inactive. 
A Monte Carlo time step consists of $N(t)$ such moves, where $N(t)$ is the number of active sites at the beginning of the time step. 
Note that, although the time elapsed is measured in a simplified way with respect to an exact simulation of the continuous-time process, it is asymptotically correct for long times. 

We have performed seed simulations \cite{GdlT}, in which, initially, all but the central site were inactive. Performing typically $10^6$ independent runs we measured the probability $P(t)$ that the system has not yet trapped in the absorbing state up to time $t$. When studying the time-dependent behavior of $P(t)$, the length of the arms was chosen to be large enough so that the end sites of arms have never been activated. 
The long time limit of the survival probability (in the infinite system) 
$P=\lim_{t\to\infty}P(t)$ serves as a local order parameter of the phase transition, which occurs at the bulk critical point $\lambda_c=3.29785(2)$ of the one-dimensional model \cite{jd}.

%%%%%%%%%%%%%%%%%%%%%%%%%%%%%%%%%%%%%%%%%%%%%%%%%%%%%%%%%%%%%%%%%%%%%%%%%%%
\section{Numerical results}
\label{sec:numerical} 

\begin{figure}[ht]
\includegraphics[width=8cm]{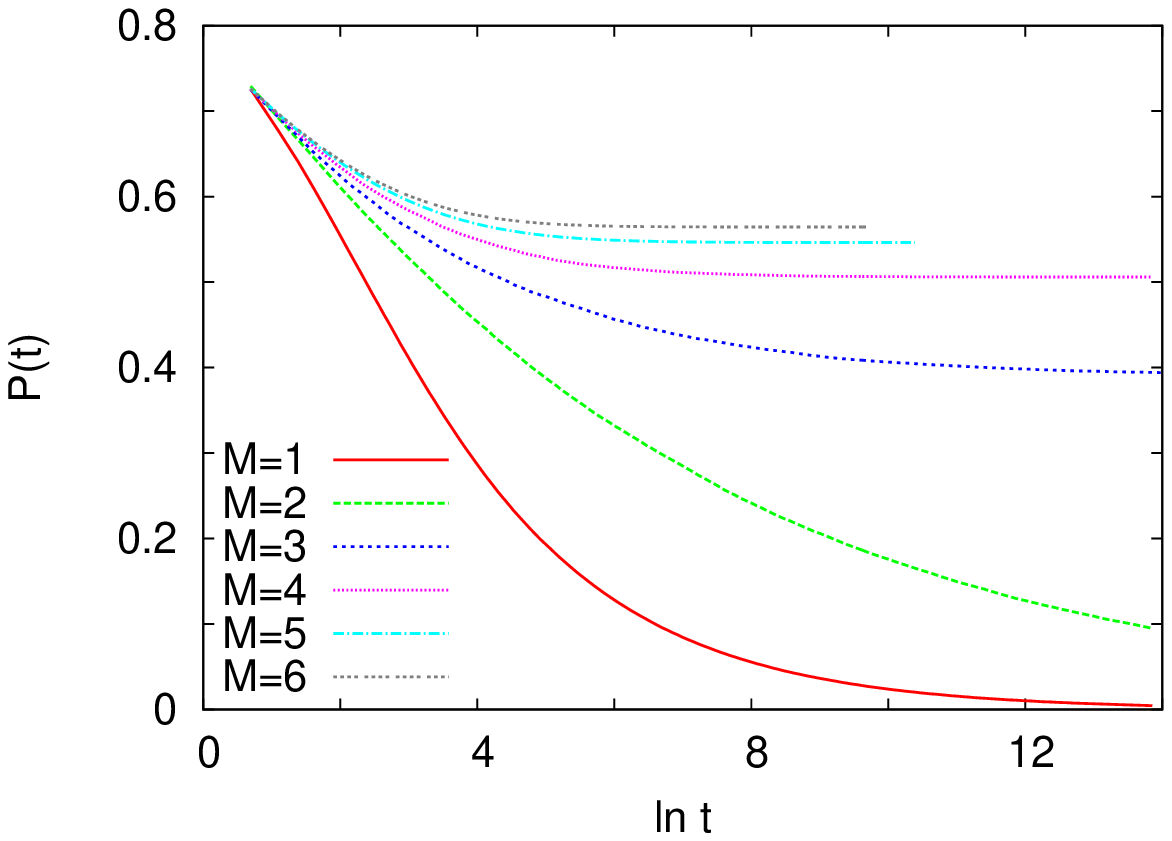} 
\includegraphics[width=8cm]{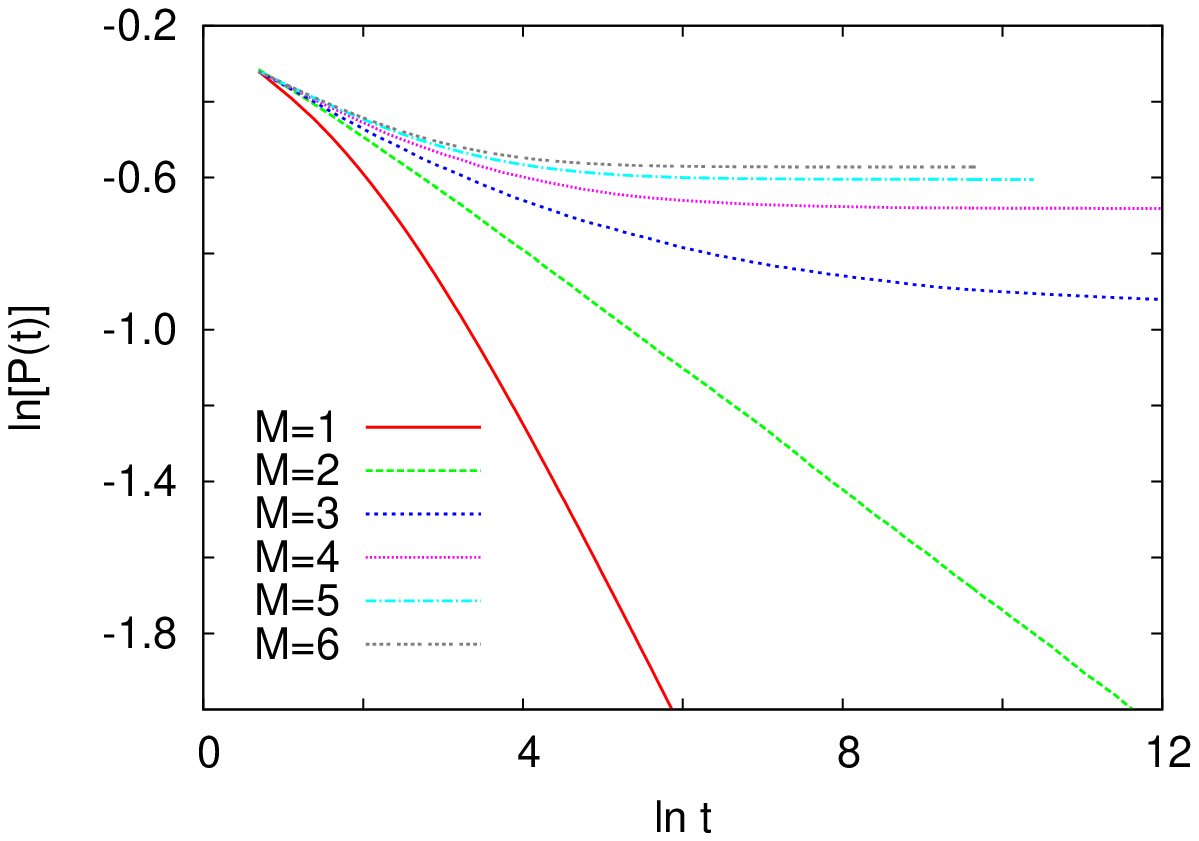} 
%pt.eps
%ptlog.eps
\caption{\label{fig_pt} (Color online) Top: Time-dependence of the survival probability measured in numerical simulations for different M-fold junctions with $M=1,2,\dots,6$ (from bottom to top) in the bulk critical point $\lambda=\lambda_c$. Bottom: The logarithm of the same function.  
}
\end{figure}

\begin{figure}[ht]
\includegraphics[width=8cm]{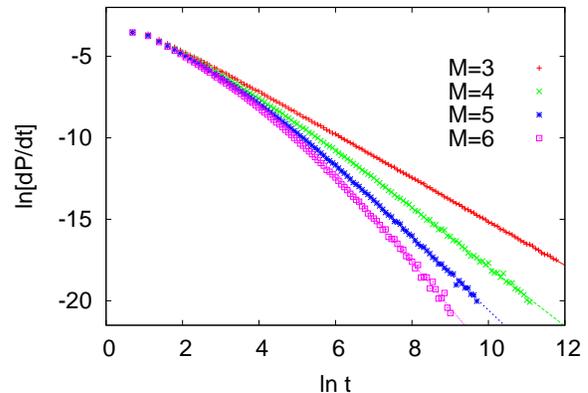}
%pder.eps  
\caption{\label{fig_pder} (Color online) The logarithm of the discrete time-derivative of the numerically obtained survival probability in the bulk critical point plotted against $\ln t$ for $M=3,4,5,6$ (from top to bottom). The straight lines are linear fits to the data, whose slope is $-\delta^{\prime}_M-1$.  
}
\end{figure}

First, we have measured the survival probability in the bulk critical point as a function of time for $M$-fold junctions with $M=1,2,\dots,6$. As can be seen in Fig. \ref{fig_pt}, for $M=1$ and $M=2$, the survival probability tends to zero in the long-time limit, in agreement with the well-known result that the critical contact process is non-surviving, i.e. the transition is continuous.  
For $M>2$, however, $P(t)$ seems to tend to a positive constant $P_c(M)$, which is an increasing function of $M$, signaling a discontinuous transition. 
The critical decay exponent $\delta_M$ of the survival probability \cite{md,odor,hhl} defined for continuous transitions as $P(t)\sim t^{-\delta_M}$ is thus formally zero here. 
%%%%%%%%%%%%%%%%%%%%%%%%%%%%%%%%%%%%%%%%%%%%%%%%%%%%%%%%%%%%%%%%%%%%%%%%%
%%%%%%%%%%%%%%%%%%%%%%%%%%%%%%%%%%%%%%%%%%%%%%%%%%%%%%%%%%%%%%%%%%%%%%%%
Inspecting the (discrete) time-derivative of $P(t)$ in a double-logarithmic plot, as shown 
in Fig. \ref{fig_pder}, it turns out to approach the limiting value according to a power law
\be 
P(t)-P_c\sim t^{-\delta^{\prime}_M},
\label{Pt_crit} 
\ee
with some $M$-dependent exponents $\delta^{\prime}_M$, which have been estimated  from linear fits to the data in Fig.  \ref{fig_pder} and are listed in Table \ref{table}. 
%%%%%%%%%%%%%%%%%%%%%%%%%%%%%%%%%%%%%%%%%%%%%%%%%%%%%%%%%%%%%%%%%%%%%%%%%
%%%%%%%%%%%%%%%%%%%%%%%%%%%%%%%%%%%%%%%%%%%%%%%%%%%%%%%%%%%%%%%%%%%%%%%%
%%%%%%%%%%%%%%%%%%%%%%%%%%%%%%%%%%%%%%%%%%%%%%%%%%%%%%%%%%%%%%%%%%%
\begin{table}[h]
\begin{center}
\begin{tabular}{|r|r|r|r|}
\hline  $M$ & $P_c$ & $\delta^{\prime}_M$ &  $\nu_{\parallel,M}$  \\
\hline  3   & 0.391(2) & 0.34(1)   &  2.20(3)           \\
\hline  4   & 0.507(2) & 0.81(3)   &  3.00(10)           \\
\hline  5   & 0.546(2) & 1.25(10)  &  3.7(1)            \\
\hline  6   & 0.564(2) & 1.8(1)    &  4.5(1)            \\
\hline
\end{tabular}
\end{center}
\caption{\label{table} Local order parameter at the critical point and critical exponents estimated by numerical simulations for different $M$-fold junctions.}
\end{table}
%%%%%%%%%%%%%%%%%%%%%%%%%%%%%%%%%%%%%%%%%%%%%%%%%%%%%%%%%%%%%%%%%%%%%%

\begin{figure}[h]
\includegraphics[width=8cm]{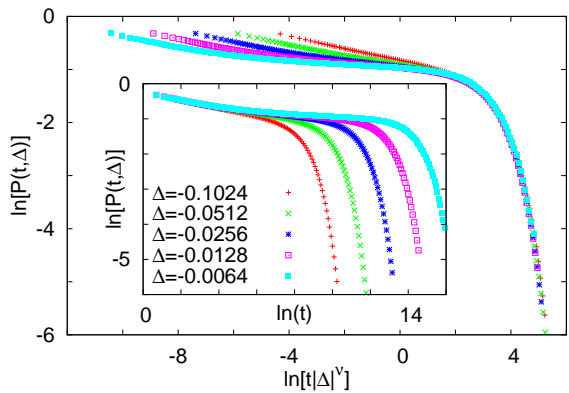} 
%nu3aa.eps  
\caption{\label{fig_nu3a} (Color online) Scaling plot of the survival probability obtained by numerical simulations for the triple junction ($M=3$) for different distances $\Delta\equiv\lambda-\lambda_c$ from the critical point, in the inactive phase ($\Delta<0$). The optimal data collapse is achieved by the value $\nu_{\parallel}=2.20$. The unscaled data are shown in the inset.    
}
\end{figure}

Next, we investigated the behavior of the temporal correlation length as the critical point is approached in the inactive phase. In order to do this, we measured the survival probability for different distances $\Delta\equiv\lambda-\lambda_c<0$ from the critical point. As can be seen in the inset of Fig. \ref{fig_nu3a}, a cutoff appears, which is shifted toward longer times as $|\Delta|$ is decreased, signaling a diverging temporal correlation length according to 
\be  
\xi_{\parallel}\sim |\Delta|^{-\nu_{\parallel,M}}.
\label{xi}
\ee  
Close to the critical point and for long times, we assume the usual scaling form of the survival probability 
\be 
P(t,\Delta)=t^{-\delta_M}f(\Delta t^{1/\nu_{\parallel,M}}),
\label{sc_inactive}
\ee
where $f(x)$ is some scaling function \cite{md,hhl}, to hold with $\delta_M=0$.
As it is illustrated for $M=3$ in Fig. \ref{fig_nu3a}, a data collapse can indeed be achieved using the scaling variable $t|\Delta|^{\nu_{\parallel,M}}$ with 
$\nu_{\parallel,3}=2.20(3)$, which is significantly higher than the corresponding bulk exponent $\nu_{\parallel}$, see in Table \ref{table_exponents}. Estimates of $\nu_{\parallel,M}$ determined in this way for further $M$-fold junctions can be found in Table \ref{table}. 
%%%%%%%%%%%%%%%%%%%%%%%%%%%%%%%%%%%%%%%%%%%%%%%%%%%%%%%%%%%%%%%%%%%%%%%%%
%%%%%%%%%%%%%%%%%%%%%%%%%%%%%%%%%%%%%%%%%%%%%%%%%%%%%%%%%%%%%%%%%%%%%%%%

\begin{figure}[h]
\includegraphics[width=8cm]{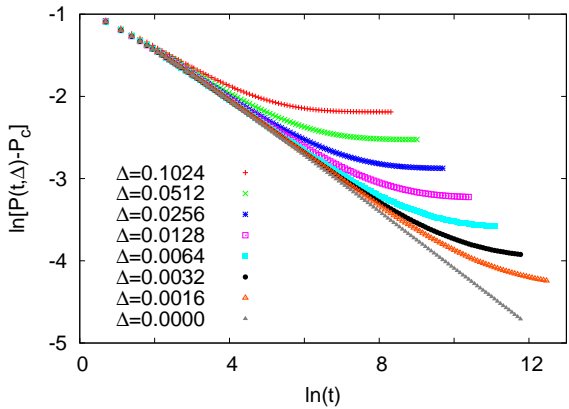}  
%nu3b0.eps
\caption{\label{fig_nu3b0} (Color online) The survival probability obtained by numerical simulations for the triple junction ($M=3$) for different values of $\Delta$ in the active phase and at the critical point (from top to bottom for decreasing $\Delta$). 
}
\end{figure}

\begin{figure}[h]
\includegraphics[width=8cm]{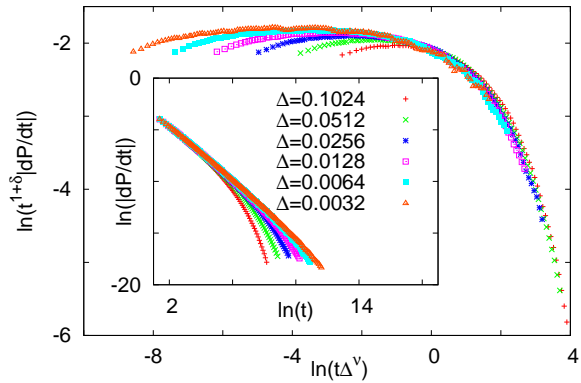}  
%nu3bb.eps
\caption{\label{fig_nu3b} (Color online) Scaling plot of the time-derivative of the survival probability obtained by numerical simulations for the triple junction ($M=3$) for different distances $\Delta\equiv\lambda-\lambda_c$ from the critical point in the active phase ($\Delta>0$). The values $\delta=0.34$ and  $\nu_{\parallel}=1.733847$ have been used. The unscaled data are shown in the inset.    
}
\end{figure}

We measured the survival probability close to the critical point in the active phase, as well. The data for the triple junction are shown in Fig. \ref{fig_nu3b0}. 
%%%%%%%%%%%%%%%%%%%%%%%%%%%%%%%%%%%%%%%%%%%%%%%%%%%%%%%%%%%%%%%%%%%%%%%%%
%%%%%%%%%%%%%%%%%%%%%%%%%%%%%%%%%%%%%%%%%%%%%%%%%%%%%%%%%%%%%%%%%%%%%%%%
Here, the deviation of $P(t,\Delta)$ from the constant $P_c$ is expected to have the scaling property
\be 
P(t,\Delta)-P_c=t^{-\delta_M^{\prime}}\tilde P(\Delta t^{1/\nu_{\parallel,M}^{\prime}}), 
\label{sc_active}
\ee
where the scaling function $\tilde P(y)$ behaves as $\tilde P(y)\sim y^{\beta_M^{\prime}}$ for $y\to\infty$. 
The exponents appearing here are related as 
\be 
\delta_M^{\prime}=\beta_M^{\prime}/\nu_{\parallel,M}^{\prime}.
\label{delta_beta}
\ee  
As can be seen in Fig. \ref{fig_nu3b0}, the local slope of the critical curve deviates significantly from the asymptotic estimate for moderate times, i.e. the corrections to scaling are expected to be strong. Therefore, using the asymptotic $\delta_M^{\prime}$, the scaling collapse according to Eq. \ref{sc_active} with the available data will be poor. 
Another difficulty is that Eq. \ref{sc_active} contains one unknown parameter ($P_c$) more than Eq. (\ref{sc_inactive}) does, and, in addition to this, the quality of the scaling is very sensitive to the error of $P_c$ for small $\Delta$ and long times.  Therefore we tried an alternative way for determining the critical exponent $\nu_{\parallel,M}^{\prime}$ and considered, namely, the time-derivative of $P(t,\Delta)$, which scales as 
\be 
\frac{\partial P(t,\Delta)}{\partial t}=t^{-\delta_M^{\prime}-1}g(\Delta t^{1/\nu_{\parallel,M}^{\prime}}), 
\label{sc_deriv}
\ee
where $g(x)$ is another scaling function. 
As it is shown in Fig. \ref{fig_nu3b} for $M=3$, a satisfactory scaling collapse is obtained by using the previous estimate $\delta_M^{\prime}=0.34$ and the bulk correlation length exponent $\nu_{\parallel,M}^{\prime}=\nu_{\parallel}$ in Table \ref{table_exponents}. 
Although the optimal scaling collapse of the data is realized with a somewhat lower value 
$\nu_{\parallel,M}^{\prime}=1.65(5)$, we conjecture that $\nu_{\parallel,M}^{\prime}$ is given by the bulk value, and the deviation may be attributed to corrections to scaling. In order to see a scaling collapse of better quality, data with smaller $\Delta$ (and correspondingly longer times) would be needed, which is beyond our present computational possibilities.  
%%%%%%%%%%%%%%%%%%%%%%%%%%%%%%%%%%%%%%%%%%%%%%%%%%%%%%%%%%%%%%%%%%%%%%%%%
%%%%%%%%%%%%%%%%%%%%%%%%%%%%%%%%%%%%%%%%%%%%%%%%%%%%%%%%%%%%%%%%%%%%%%%%

\begin{figure}[h]
\includegraphics[width=8cm]{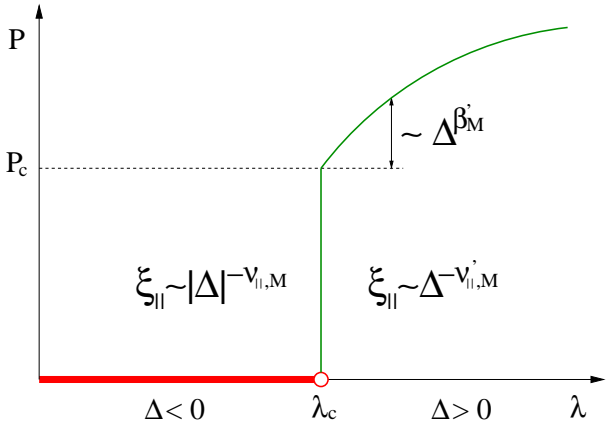}  
%op2.eps
\caption{\label{op2} (Color online) Illustration of the behavior of the local order parameter close to the transition point, together with the definition of the different singular quantities. We indicate by a red thick line the region in which the local critical behavior is influenced by the presence of a dangerous irrelevant scaling variable, see the text.    
}
\end{figure}

The scaling form in Eq. (\ref{sc_active}) implies that the order parameter in the active phase, i.e. the survival probability in the stationary state,  
$P(\Delta)\equiv \lim_{t\to\infty}P(t,\Delta)$ approaches the critical value in a singular way as 
\be 
P(\Delta)-P_c\sim \Delta^{\beta_M^{\prime}}.
\label{sc_delta}
\ee
Owing to the strong corrections mentioned above, a direct estimation of $\beta_M^{\prime}$ according to Eq. (\ref{sc_delta}) and using the numerical data in hand would be considerably below the presumably correct value. 
Instead, an indirect estimate from the data for $\delta_M^{\prime}$ using the relation in Eq. (\ref{delta_beta}) is expected to be more reliable. A schematic picture of the behavior of the local order parameter is shown in Fig.\ref{op2}.
 %%%%%%%%%%%%%%%%%%%%%%%%%%%%%%%%%%%%%%%%%%%%%%%%%%%%%%%%%%%%%%%%%%%%%%%%%
%%%%%%%%%%%%%%%%%%%%%%%%%%%%%%%%%%%%%%%%%%%%%%%%%%%%%%%%%%%%%%%%%%%%%%%%

%%%%%%%%%%%%%%%%%%%%%%%%%%%%%%%%%%%%%%%%%%%%%%%%%%%%%%%%%%%%%%%%%%%%%%%%%%%
\section{Scaling considerations}
\label{sec:scaling}

The properties of the phase transition of the contact process at multiple junctions have unusual features: it is of mixed order and there is a correlation-length exponent asymmetry at the two sides of the transition point. Similar type of critical behavior has already been observed for the Ising model in the same geometry \cite{itb,cardy,grassberger} and it has been explained in terms of irrelevant scaling variables \cite{it}. In the following, we generalize this scaling theory for the dynamical critical behavior of the contact process.

Here we use the well-known mapping \cite{cardy_book}, which relates the dynamics of a $d$-dimensional system to the static behavior of a $(d+1)$-dimensional system, where scaling in the extra dimension is anisotropic, characterized by an anisotropy exponent $z>1$. In the case of the $d=1$ contact process, we have a $(1+1)$-dimensional static problem, where coordinates in the space direction will be denoted by $x$, the correlation length by $\xi_{\perp}$, while, in the extra (time) direction, we use the notations $t$ and $\xi_{\parallel}$, respectively.
In this static problem $\xi_{\perp}\sim \xi_{\parallel}^z$ with $z\ge 1$, and we write the local free energy per degree of freedom at the junction $f(\Delta,h_1,\Delta_1,t)$ as the function of the bulk and the junction control parameters, $\Delta$ and $\Delta_1$, respectively, as well as of a local ordering field, $h_1$. (We have omitted here the bulk ordering field, $h$.) According to scaling theory, when lengths are rescaled by a factor $b>1$, so that $x \to x/b$, then the free-energy density satisfies the relation:
\be
f(\Delta,h_1,\Delta_1,t)=b^{-z}f(\Delta b^{1/\nu_{\perp}},h_1 b^{y_{h_1}},\Delta_1 b^{y_{\Delta_1}},t/b^z)\;,
\label{f_scal}
\ee
and, similarly, we have for the parallel correlation length:
\be
\xi_{\parallel}(\Delta,h_1,\Delta_1,t)=b^{z}\xi_{\parallel}(\Delta b^{1/\nu_{\perp}},h_1 b^{y_{h_1}},\Delta_1 b^{y_{\Delta_1}},t/b^z)\;.
\ee
(Note that, due to anisotropic scaling, one should define pairs of scaling exponents: $y_{h_1}^{\perp}$, $y_{h_1}^{\parallel}$ and
$y_{\Delta_1}^{\perp}$, $y_{\Delta_1}^{\parallel}$. Here and in the following $y_{h_1}$ and $y_{\Delta_1}$ will correspond to $y_{h_1}^{\perp}$ and $y_{\Delta_1}^{\perp}$, respectively.)

The local order parameter, which is analogous to the survival probability in the contact process, is related to the local free energy density as $P=\partial f/\partial h_1$. This scales as:
\be
P(\Delta,h_1,\Delta_1,t)=b^{-z+y_{h_1}}f(\Delta b^{1/\nu_{\perp}},h_1 b^{y_{h_1}},\Delta_1 b^{y_{\Delta_1}},t/b^z)\;
\ee
and, due to ordering at the junction, $P$ is scale-independent, thus we have
\be
y_{h_1}=z\;.
\ee

The local scaling field associated with the local control parameter $\Delta_1$ is an irrelevant variable, thus $y_{\Delta_1}<0$. In the following, we assume, as in the isotropic case\cite{it}, that $\Delta_1$ is a harmless irrelevant scaling variable for $\Delta \ge 0$, whereas it is dangerous irrelevant in the inactive phase, $\Delta < 0$.

For $\Delta \ge 0$, the scaling functions are analytic in $\Delta_1$ and can be expanded in Taylor series. We obtain in leading order for the local order parameter:
\be
P(\Delta,h_1,\Delta_1,t)-P_c=b^{y_{\Delta_1}}{\tilde P}(\Delta b^{1/\nu_{\perp}},h_1 b^{y_{h_1}},t/b^z)\;,
\ee
which, using $b=t^{1/z}$ at the critical point $\Delta=h_1=0$, scales as:
\be
P(t)-P_c \sim t^{y_{\Delta_1}/z}\;.
\ee
Thus comparing with Eq.(\ref{Pt_crit}), we obtain: 
\be
\delta^{\prime}_M=-y_{\Delta_1}/z\;.
\ee
Similarly, with $b=\Delta^{-\nu_{\perp}}$,
we obtain the relation 
\be
\beta_M^{\prime}=-y_{\Delta_1}\nu_{\perp}\;.
\ee
For $\Delta<0$, when the scaling functions are non-analytic in $\Delta_1$, we generalize the functional form of the correlation length for isotropic systems\cite{it} as follows:
\be
\xi_{\parallel}(\Delta,h_1,\Delta_1,x)=\Delta_1^{-\epsilon} {\tilde \xi}_{\parallel}(\Delta,h_1 \Delta_1^{-\epsilon},x)\;,
\ee
which has the scaling relation:
\be
\xi_{\parallel}(\Delta,h_1,\Delta_1,x)=b^{z-{\epsilon}y_{\Delta_1}} {\tilde \xi}_{\parallel}(\Delta b^{1/\nu_{\perp}},h_1 b^{z-{\epsilon}y_{\Delta_1}},x/b)\;.
\ee
Now taking $b=x=\xi_{\perp}$, we obtain at the critical point $\xi_{\parallel} \sim \xi_{\perp}^{z_M}$, with the anisotropy exponent at the junction in the inactive phase: $z_M=z-{\epsilon}y_{\Delta_1}$. Thus
the parallel correlation length exponent can be expressed as:
\be
\nu_{\parallel,M}=\nu_{\parallel}-\epsilon y_{\Delta_1}\nu_{\perp}=\nu_{\parallel}(1+\epsilon \delta^{\prime}_M)\;,
\label{nu_relation}
\ee
which constitutes a relation between the measured exponents, $\nu_{\parallel,M}$ and $\delta^{\prime}_M$ and the exponent in the irrelevant scaling combination, $\epsilon$. 

For the Ising model ($z=1$), where $\beta_1=1/2$ and $\nu_{\parallel}=1$, the critical exponents at the junction can be exactly expressed in terms of $\beta_1/\nu_{\parallel}$ and $M$ as follows \cite{it}:
\beqn
\nu_{\parallel,M}&=&M \beta_1 \cr
\delta^{\prime}_M&=&(M-\nu_{\parallel}/\beta_1)/2\;.
\label{final_rel}
\eeqn
These imply, through Eq.(\ref{nu_relation}), that $\epsilon$ is independent of $M$: 
\be
\epsilon=2 \beta_1/\nu_{\parallel}\;.
\label{epsilon}
\ee
Using Eq. (\ref{nu_relation}) and the exponents measured for the contact process and given in Table \ref{table}, the calculated $\epsilon=(\nu_{\parallel,M}/\nu_{\parallel}-1)/\delta^{\prime}_M$ has only a weak variation with $M$. 
Furthermore, the relations in Eqs. (\ref{final_rel}) seem to be satisfied by the numerical values in Table \ref{table}, at least for $M>3$ within the error of the estimation. For $M=3$, the exponent $\delta^{\prime}_3$ is somewhat above the value calculated from Eqs. (\ref{final_rel}). 
In spite of this slight discrepancy with the numerical data for $M=3$, we cannot exclude the validity of relations in Eqs. (\ref{final_rel}) for the contact process with certainty, as slow corrections may be present in the numerical data, which are hard to detect on the time scales available by the simulations. 
Having the conjectured values of the scaling exponents, we can obtain the singular behavior of other observables via Eq.(\ref{f_scal}), as well.

%%%%%%%%%%%%%%%%%%%%%%%%%%%%%%%%%%%%%%%%%%%%%%%%%%%%%%%%%%%%%%%%%%%%%%%%%%%
\section{The effect of quenched disorder}
\label{sec:disorder} 

\begin{figure}[h]
\includegraphics[width=8cm]{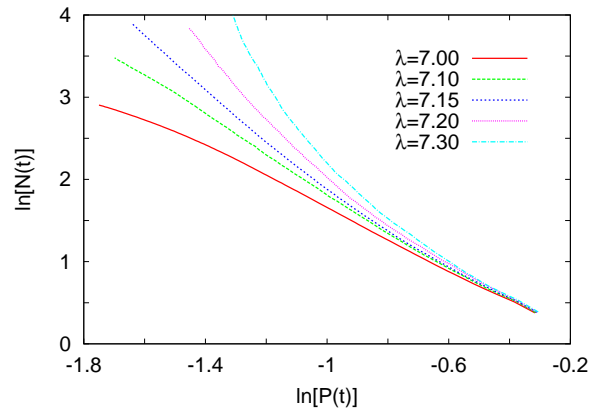}  
%npq2.eps
\caption{\label{npq2} (Color online) The logarithm of the average number of active sites plotted against the logarithm of the survival probability in the disordered contact process in one dimension for different values of the control parameter $\lambda$, from bottom to top for increasing $\lambda$. The critical point is estimated to be at $\lambda_c=7.15$.     
}
\end{figure}

\begin{figure}[h]
\includegraphics[width=8cm]{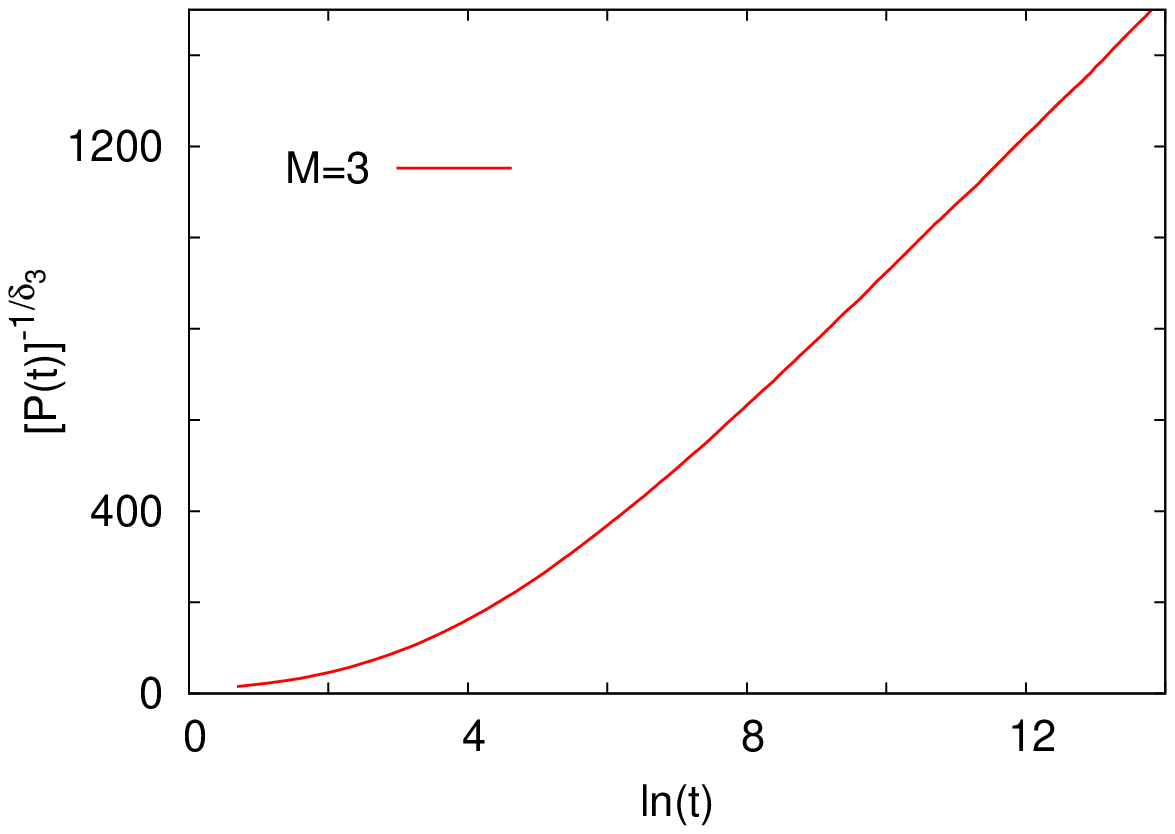}
\includegraphics[width=8cm]{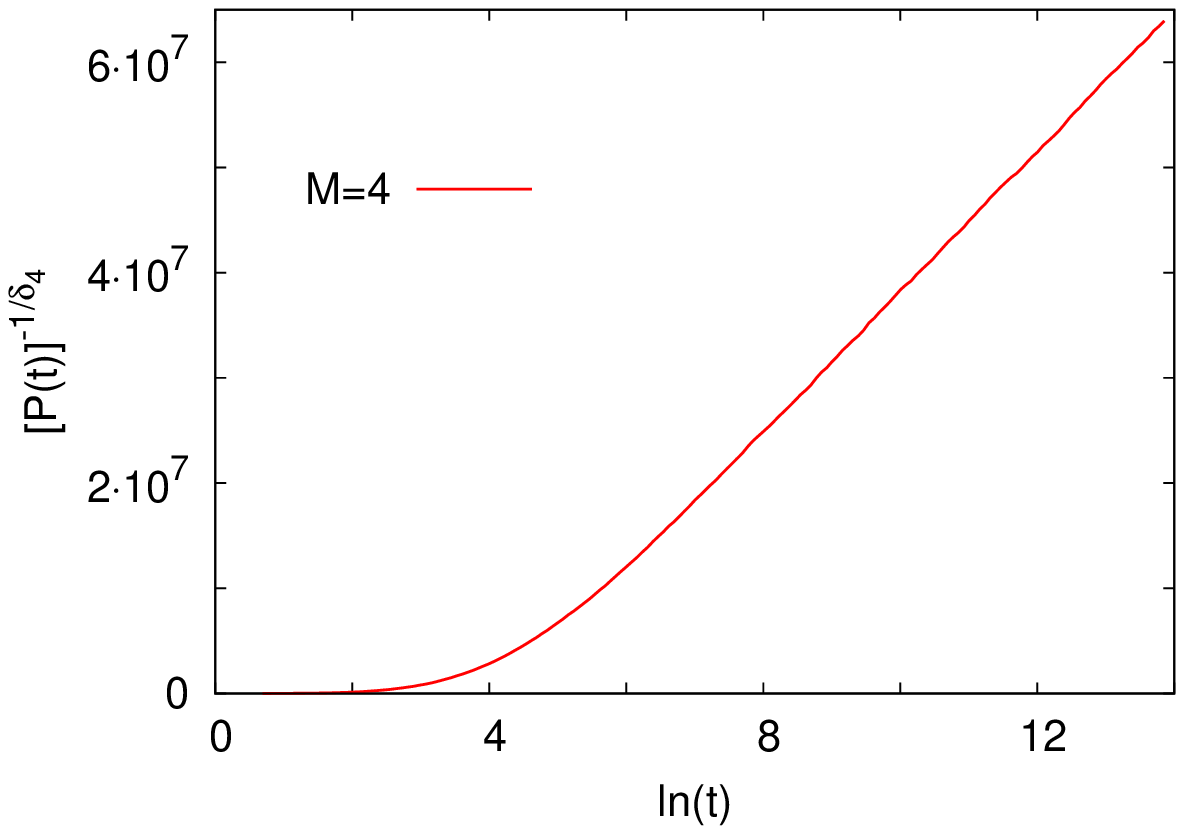}  
%pq3.eps
%pq4.eps

\caption{\label{pq} (Color online) Time-dependence of the survival probability  measured in numerical simulations for the disordered contact process at a triple (top) and a quadruple (bottom) junction at the bulk critical point $\lambda_c=7.15$. The exponents $\overline{\delta}_3=0.139194$ and 
$\overline{\delta}_4=0.043967$ have been taken from Ref. \cite{qcpmj}. According to Eq. (\ref{pt_qcp}), the dependence as plotted in the figures must be linear if the correct exponents are used.     
}
\end{figure}

In the presence of quenched random transition rates, the one-dimensional contact process displays a continuous phase transition. The scaling laws close to the critical point, which can be determined by a strong-disorder renormalization group (SDRG) method \cite{hiv} and are believed to be valid at least for strong disorder, are much different from those of the homogeneous system. The striking difference compared to the scaling laws of pure systems is that they contain the logarithm of time rather than the time itself. For instance, the survival probability follows the scaling form
\be 
P(\ln t,\Delta)=(\ln t)^{-x_2/\psi}\tilde P(\Delta(\ln t)^{1/(\psi\nu_{\perp})}),
\label{sc_qcp}
\ee 
where the scaling function behaves as $\tilde P(y)\sim y^{x_2\nu_{\perp}}$ for $y\to\infty$. 
The critical exponents appearing here are $\psi=1/2$, $\nu_{\perp}=2$, and $x_2=(1+\sqrt{5})/4$ \cite{hiv}. 
In the semi-infinite system, a similar scaling form as Eq. (\ref{sc_qcp}) is valid with the same $\psi$ and $\nu_{\perp}$ as in the bulk, but a different scaling dimension of the local order parameter, $x_1=1/2$. 
A recent SDRG study of the model near an $M$-fold junction showed that the phase transition remains continuous for any value of $M$ \cite{qcpmj}. 
Thus, quenched disorder makes the originally discontinuous phase transition of the clean system for $M>2$ continuous, which is not an uncommon phenomenon for systems with first-order transitions. 
The scaling dimension $x_M$ of the order parameter is a rapidly decreasing function of $M$, and can be obtained as the smallest eigenvalue of an $M\times M$ matrix \cite{qcpmj}. The numerical values for $M=3$ and $M=4$ are 
$x_3=0.06959\dots$ and $x_4=0.02198\dots$, respectively. 
These outcomes of the SDRG method have not been confirmed by simulations so far, so we aimed at performing Monte Carlo simulations of the disordered system and study its critical scaling. 
Quenched disorder was realized by using site-dependent deactivation and activation rates, $1/(1+\lambda_i)$ and $\lambda_i/(1+\lambda_i)$, respectively. 
Here, $\lambda_i\equiv r_i\lambda$, where $r_i$ is a quenched i.i.d. random parameter taking values $1$ and $0.2$ with probabilities $1/2$. 
Performing seed simulations in typically $10^5-10^6$ random samples (with one run per sample), the survival probability $P(t)$ and the average number of active sites $N(t)$ have been measured. 
First, we determined the location of the critical point by plotting $\ln N(t)$ against $\ln P(t)$ and looking for a $\lambda$, for which the dependence is asymptotically linear \cite{vd}, see Fig. \ref{npq2}.   
%%%%%%%%%%%%%%%%%%%%%%%%%%%%%%%%%%%%%%%%%%%%%%%%%%%%%%%%%%%%%%%%%%%%%%%%%
%%%%%%%%%%%%%%%%%%%%%%%%%%%%%%%%%%%%%%%%%%%%%%%%%%%%%%%%%%%%%%%%%%%%%%%%
This gives the estimate $\lambda_c=7.15(5)$.
Next, we measured the dependence of the survival probability on time in seed simulations for triple and quadruple junctions at $\lambda=\lambda_c$. 
At the critical point, $P(t)$ decays according to Eq. (\ref{sc_qcp}) as 
\be
P(t)\sim (\ln t)^{-\overline{\delta}_M},
\label{pt_qcp}
\ee
where $\overline{\delta}_M=x_M/\psi$. Using the numerical values of $\overline{\delta}_M$ obtained by the SDRG method, we plotted $[P(t)]^{-1/\overline{\delta}_M}$ against $\ln t$. 
As can be seen in Fig. \ref{pq}, the dependence for $M=3$ and $M=4$ is linear for long times, in agreement with Eq. (\ref{pt_qcp}), confirming thus the correctness of the exponents obtained by the SDRG method.
%%%%%%%%%%%%%%%%%%%%%%%%%%%%%%%%%%%%%%%%%%%%%%%%%%%%%%%%%%%%%%%%%%%%%%%%%
%%%%%%%%%%%%%%%%%%%%%%%%%%%%%%%%%%%%%%%%%%%%%%%%%%%%%%%%%%%%%%%%%%%%%%%%

%%%%%%%%%%%%%%%%%%%%%%%%%%%%%%%%%%%%%%%%%%%%%%%%%%%%%%%%%%%%%%%%%%%%%%%%%%%
\section{Discussion}
\label{sec:discussion}

We have studied in this paper the phase transition of the contact process near $M$-fold junctions by numerical simulations and scaling considerations. We have found that, for $M>2$, the local order parameter displays a discontinuity at the critical point, as opposed to the translationally invariant ($M=2$) and semi-infinite system ($M=1$), where the transition is continuous. 
The ordering of the system near the junction is a remarkable feature, as a local supercritical creation rate in an otherwise critical system is not able to induce ordering --- not even the local critical exponents are modified, only an additive correction appears \cite{bh}.
The temporal correlation length is found to diverge algebraically with the distance from the transition point. The corresponding exponents are, however, different on the two sides of the transition; in the active phase it is found to be close to the bulk value, while, in the inactive phase, it exceeds the bulk value and increases roughly linearly with $M$.   
The model thus provides an example for mixed-order transitions and that, for the subclass where the correlation length diverges as a power law. (An alternative is the Kosterlitz-Thouless type essential singularity.)
A difference compared to other models is that, here, a local order parameter of  a translationally non-invariant model is concerned. 
Nevertheless, the critical behavior observed here is not restricted to the immediate vicinity of the junction. The local order parameter remains discontinuous in any large distance $l$ from the junction and its critical value decays asymptotically as 
$P_c(l)\sim l^{-\beta/\nu_{\perp}}$\footnote{This is similar to the decay of the density profile in the one-dimensional model with an active wall \cite{hhl,fd}.}. 
Obviously, in a distance $l$ from the junction, bulk scaling laws can be observed within the crossover length scale $l$ or, in case of time-dependence, within the time scale $l^z$, where $z=\nu_{\parallel}/\nu_{\perp}$ is the bulk dynamical exponent. But beyond this time scale, a crossover to scaling laws found at the junction occurs.  
We have also demonstrated by numerical simulations that quenched spatial disorder makes the transition continuous confirming thereby earlier results obtained by the SDRG method.  

The behavior found for the contact process is analogous to that of the phase transition of the two-dimensional Ising model near $M$-fold junctions \cite{itb,cardy,itb,grassberger}. 
In that model, the transition is continuous for $M\le 2$ but, for $M>2$, the local magnetization is finite in the critical point. 
A further similarity is that the temporal correlation length exponent is asymmetric. In the ferromagnetic phase, it is identical to the bulk value ($\nu_{\parallel}^{\prime}=1$), but in the paramagnetic phase it is different and given by $\nu_{\parallel,M}=M/2$ for $M>2$.   
For that model and other models showing a first-order local transition, a general scaling theory has been developed, which explains the anisotropic scaling behavior and the exponent asymmetry by that the local control parameter is a dangerous irrelevant variable \cite{it}.
This type of scaling theory has been generalized in this paper for systems having an anisotropy exponent, $z>1$. Applying this scaling approach for the contact process, we could explain the observed anomalous local critical behavior. 
Furthermore, we have given a conjecture for the values of the local scaling exponents in terms of $\beta_1/\nu_{\parallel}$ and $M$, in Eqs.(\ref{final_rel}). 
But, to judge the validity of these conjectures with a larger certainty, more accurate numerical data are needed. 

Another well-known model that shows a discontinuous surface transition at the continuous bulk transition point is, the Hilhorst-van Leeuwen model \cite{HvL}. 
This is a two-dimensional Ising model (or, equivalently, a one-dimensional transverse-field Ising chain), where the local control parameter decays with the distance $l$ from a free surface as $\Delta_l\simeq Al^{-1/\nu_{\perp}}$. 
Depending on $A$, the surface transition can be continuous or discontinuous (of mixed-order) with critical exponents varying with $A$. The scaling theory of Ref. \cite{it} applies also to this model. 
Based on the close analogy between the Ising model and contact process near multiple junctions, a similar scenario is expected also for the contact process with a Hilhorst-van Leeuwen type extended defect. The numerical study of this problem is in progress and the preliminary results indeed confirm this expectation. 

An interesting question is whether the phenomenon that a multiple junction induces a discontinuity for models with a continuous bulk transition is a general rule and, if not, under which condition it occurs.
A further direction in which the findings of the present work could be generalized is the increasing dimensionality. We considered the contact process on chains connected to a common site, but one can also imagine semi-infinite planes attached to each other at a common line, just as for the classical formulation of the related problem of the Ising model. In this case, we expect a discontinuous phase transition near the common line, analogously to the one-dimension lower problem. 
The study of the above questions is deferred to future research.

\begin{acknowledgments}
We thank Peter Grassberger for helpful correspondence.
This work was supported by the Hungarian Scientific Research Fund under Grants
No. K109577 and No. K115959. 
\end{acknowledgments}

%%%%%%%%%%%%%%%%%%%%%%%%%%%%%%%%%%%%%%%%%%%%%%%%%%%%%%%%%%%%%%%%%%%%%%%%%%%
%%%%%%%%%%%%%%%%%%%%%%%%%%%%%%%%%%%%%%%%%%%%%%%%%%%%%%%%%%%%%%%%%%%%%%%%%%%

\end{document}